\documentclass[reprint,aps,pra,nofootinbib]{revtex4-2}
\usepackage{graphicx}
\usepackage{amsmath,amssymb}
\usepackage{epstopdf}
\usepackage{epstopdf}
\usepackage{float}
\usepackage{textcomp}
\usepackage{color}
\usepackage{siunitx}
\usepackage{float}
\usepackage{ulem}
\usepackage{bbm}

\newcommand{\ket}[1]{|#1\rangle}

\DeclareSIUnit\gauss{G} 
\usepackage{pgf}

\begin{document}

\title[Atoms in lowest Landau level]{Chiral edge dynamics and quantum Hall physics in synthetic dimensions with an atomic erbium Bose-Einstein condensate}
\author{Roberto Vittorio Roell,$^{*,\dagger}$ Arif Warsi Laskar,$^{*,\ddagger}$ Franz Richard Huybrechts, and Martin Weitz}
\affiliation{Institute of Applied Physics, University of Bonn, 53115 Bonn, Germany}
\date{\today}

\def\thefootnote{$\dagger$}\footnotetext{roell@iap.uni-bonn.de}
\def\thefootnote{$\ddagger$}\footnotetext{laskar@iap.uni-bonn.de}
\def\thefootnote{*}\footnotetext{These two authors contributed equally}

\begin{abstract}
	
	Quantum Hall physics is at the heart of research on both matter and artificial systems, such as cold atomic gases, with non-trivial topological order. We report on the observation of a chiral edge current by transferring atomic wavepackets simultaneously to opposite edges of a synthetic Hall system realized in the two-dimensional state space formed by one spatial and one synthetic dimension encoded in the $J=6$ electronic spin of erbium atoms. To characterize the system, the Hall drift of the employed atomic Bose-Einstein condensate in the lowest Landau-like level is determined. The topological properties are verified by determining the local Chern marker, and upon performing low-lying excitations both cyclotron and skipping orbits are observed in the bulk and edges respectively. Future prospects include studies of novel topological phases in cold atom systems.

\end{abstract}


	
\date{\today}
\maketitle

For charged particles such as electrons, the use of magnetic fields is both an attractive manipulation tool on the classical level, yielding a Lorentz force, as well as in quantum physics, where for two-dimensional electron gases both the integer and the fractional quantum Hall effect have been observed \cite{klitzing80, Tsui82}. Quantum Hall systems have topologically protected edge states, and their bulk is characterized by a non-vanishing Chern number \cite{Thouless1982}. Atoms are electrically neutral, but it has been demonstrated that magnetic fields can be emulated by utilizing the Coriolis force in rotating systems~\cite{madison2000vortex,bretin2004fast,schweikhard2004rapidly,Tung06,Williams10}, phase imprinting~\cite{jaksch2003creation,gerbier2010gauge,cooper2011optical}, or lattice shaking \cite{Sorensen05,Eckardt05,Lignier07}. The interest in studying systems with non-trivial topological order with a well-controlled cold atom system arises from prospects to study a wide class of topological phases \cite{Regnault03,Ruseckas05,Kennedy13,Jotzu14,Atala2014,Lohse18,Cooper19}. Recent theory and experimental work have shown that artificial gauge fields can also be induced in the two-dimensional state space formed by one internal atomic and one external degree of freedom \cite{celi2014synthetic,Mancini2015,stuhl2015visualizing,Chalopin2020}, which builds upon earlier work on the manipulation of cold atoms with Doppler-sensitive Raman transitions showing spin-orbit coupling \cite{Spielman09,lin2009synthetic,lin2011spin,Spielman2015,Galitski2013}. For a suitable arrangement, the tensor light shift of the coupled multilevel system can compensate the quadratic dispersion of atoms, resulting in a flattening of the dispersion of the lower energy bands in the bulk of the system that is reminiscent to the Landau levels of a charged particle in a real magnetic field \cite{Chalopin2020}. From an experimental point, the use of internal Zeeman levels allows for spin-selective detection, which reveals the signature of a quantized Hall response for bosonic systems already at low filling factors. A recent experiment with a cold thermal gas of atomic dysprosium ($J = 8$) has realized a synthetic quantum Hall system with a large enough number of Zeeman levels that distinct bulk and edge behavior can be observed by studying dynamics associated with the lowest lying Landau-like bands \cite{Chalopin2020}. 

Here we report on the realization of a synthetic Hall ribbon with an atomic erbium Bose-Einstein condensate, and we describe both transport effects associated with the low-lying flattened band structure and the higher energetic, more complex band structure. The coupling of optical Raman beams with the erbium atoms creates a synthetic magnetic field in the state space spanned by one spatial and one synthetic dimension, the latter encoded in the internal Zeeman quantum number $m_J=-6,...,6$ of the atoms. The size of the synthetic dimension also in the erbium case is sufficient to allow for distinct bulk and edge behavior, as is clear from the observed data for the Hall drift of atoms in the lowest Landau-like level. The topological properties are characterized by determining the local Chern marker, and upon employing excitations to the next higher band, both skipping orbits and cyclotron motion are observed. To reveal chiral edge dynamics, a bichromatic drive with the Raman beams is applied to populate simultaneously upper and lower edges of the synthetic system such that the higher excited levels of the band structure become relevant. Our observed experimental data for the dynamics of the chiral current is understood from the known band structure of the dressed state system, with the influence of both the discretization along the synthetic dimension in terms of  the Zeeman structure and the photon recoil becoming relevant.

\begin{figure*}[htb]
	\includegraphics[width=\linewidth]{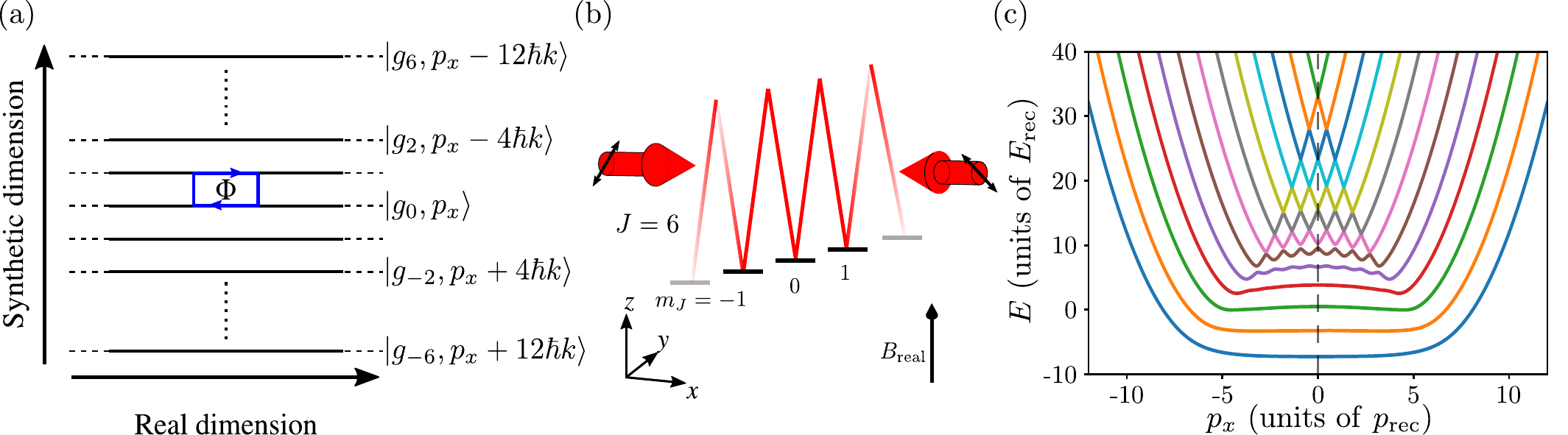}
	\caption{(a) Schematics of the synthetic Hall system realized in the state space of the internal Zeeman levels $m_J=-6, …, 6$ of erbium ground state atoms and one real space dimension. The accumulation of a geometric phase $\Phi$ along an exemplarily shown closed path is indicated. (b) Coupling scheme of atomic Zeeman levels by the two counterpropagating optical Raman beams. (c) Dispersion relation in the presence of coupling by the Raman beams for $\Omega=1.5E_{\mathrm{rec}}/\hbar$. The atom-light coupling leads to spin-orbit coupling and a flattening of low-energy bands, which exhibit similarities to Landau levels.}
	\label{fig:setup} 
\end{figure*}

A schematic of the realized synthetic quantum Hall system in state space representation is shown in Fig.~\ref{fig:setup}(a). Here, the vertical axis shows the synthetic dimension encoded in the Zeeman level structure $m_J=-6,...,6$ of the electronic ground state ($J=6$) of erbium atoms ($^{168}\mathrm{Er}$), and the horizontal axis the real dimension, directed along the Raman beams (x-axis) which are irradiated in a counterpropagating manner. The laser beams are detuned about $10^5$ linewidths to the blue of the $\mathrm{[Xe]}4f^{12}6s^2(^3H_6)-4f^{11}(^4I_{15/2})5d_{5/2}6s^2$ narrow linewidth transition ($\Gamma/2\pi\approx \SI{8}{\kilo\hertz}$) of the atomic erbium near $\SI{841}{\nano\meter}$ wavelength, and induce a Raman coupling between the states \mbox{$\ket{g_{m_J},p_x-2m_J\hbar k_x}$} and \mbox{$\ket{g_{m_J+1},p_x-2(m_J+1)\hbar k_x}$}, where $m_J$ denotes the Zeeman state and the second quantum number refers to the atomic momentum along the axis of the laser beams (x-axis) respectively. Further, $k_x\equiv k=2\pi/\lambda $ is the optical wavevector and $\lambda$ the wavelength. The coupling is such that each transition between adjacent Zeeman levels is accompanied with an atomic momentum change of two-photon recoils, and $p_x$ can be identified as the canonical momentum. Along the synthetic dimension, the Raman coupling induces a hopping term of the form $\propto e^{i\phi(x)}$, where $\phi(x)=2kx$ is the phase imprinted by the Raman beams. Along a closed loop in state space, as indicated exemplarily in Fig.~\ref{fig:setup}(a), atoms acquire a Peierls phase $\Phi=\phi(x+L)-\phi(x) = 2kL$, where $L$ denotes the displacement. This phase takes the place of the Aharanov-Bohm phase acquired by a charged particle in a real magnetic field and allows one to define the magnitude of the here produced synthetic gauge field.

In the used experimental configuration, the Raman beams are both linearly polarized along orthogonal directions, each forming an angle of $\SI{45}{\degree}$ to the z-axis [see Fig.~\ref{fig:setup}(b)] along which the magnetic bias field is oriented. The bias field causes a frequency splitting of $\omega_z/2\pi\cong\SI{307}{\kilo\hertz}$ between adjacent Zeeman sublevels. It can be shown that the dynamics of the atomic multilevel system subject to the laser beams irradiation, accounting for the Clebsch-Gordon coefficients of this $J=6 \rightarrow J’=7$ transition can be described by the Hamiltonian \cite{chalopin2019quantum} (see also the Appendix)
	
\begin{equation}
	\hat{H}_{\mathrm{a}} =\frac{(\hat{p}_x-p_{\mathrm{rec}}\hat{J}_z)^2}{2m}-\hbar\Omega\hat{J}_y - \hbar\Omega\frac{\hat{J}_z^2}{2J+3},
	\label{eq:synham}
\end{equation}
where $m$ denotes the atomic mass, $p_\mathrm{rec}=2\hbar k$ refers to the recoil momentum of a two-photon transition, and $\hat{J}_y$ and $\hat{J}_z$ are spin projection operators. Further, $\Omega$ is the Raman coupling, with typically $\Omega\cong1.5 E_\mathrm{rec}/\hbar$ for $\sim\SI{25}{\milli\watt}$ power of the Raman beams on a $\SI{280}{\micro\meter}$ beam diameter, where $E_\mathrm{rec}=(2\hbar k)^2/2m\cong h\times \SI{6.713}{\kilo\hertz}$ is the two-photon recoil energy. The first term of Eq.~(\ref{eq:synham}) is the kinetic energy from which here also spin-orbit coupling emerges, the second one expresses the coupling between Zeeman levels, and the third one the light shift, which for the used polarization configuration has tensorial form such that it for a suitable Rabi coupling allows to counteract the quadratic dispersion. Figure~\ref{fig:setup}(c) shows the resulting expected spectrum for the used experimental parameters. When approximating the lowest eigenenergies of Eq.~(\ref{eq:synham}) as equidistantly spaced flat bands, one can perform a mapping to the Landau Hamiltonian for an electron of charge $–e$ in a magnetic field $B$ directed orthogonal to the xy-plane, where $\hat{y}$ and $\hat{p}_y$ are conjugate variables and for sake of easy identification we also use $m$ for the mass

\begin{equation}
	\hat{H}_{\mathrm{L}} =\frac{(\hat{p}_x-eB\hat{y})^2}{2m}+\frac{\hat{p}_y^2}{2m},
	\label{eq:landuaham}
\end{equation}
when identifying $p_\mathrm{rec}$ with $eB$ and the spin operator $\hat{J}_z$, which points along the synthetic dimension, with $\hat{y}$ \cite{Chalopin2020}. The outer Zeeman components $m_J=\pm J$ are the edges of the synthetic Hall ribbon.

In a typical experimental run, we have prepared an erbium $(^{168}\mathrm{Er})$ Bose-Einstein condensate spin polarized in the $m_J=-6$ Zeeman state of $\sim50000$ atoms by evaporative cooling in a crossed optical dipole trap. After extinguishing the dipole trapping beams, atoms were exposed to Raman manipulation by activating the driving optical beams within a $\SI{150}{\micro\second}$ long ramp of increasing intensity. At this time, the frequency difference $\Delta\omega$ of Raman beams is set such that atoms being at rest in the lab frame and in $m_J=-6$ are loaded into the ground band, with which the used value of $\Delta\omega/2\pi\cong \SI{245}{kHz}$ corresponds to a value of $p_x \approx -10.5~p_\mathrm{rec}$ [see Eq.~(\ref{eq:synham})], i.e., in the far left of the dispersion relation, where the ground band is essentially purely given by $\ket{g_{-6},p_x+12\hbar k}$. State preparation in the ground band at a desired value of the canonical momentum $p_x$ was performed by subsequently adiabatically ramping the frequency difference of Raman beams with a rate of $\SI{1}{\kilo\hertz\per\micro\second}$ to a final value $\Delta\omega=\omega_z+2(p_x/p_\mathrm{rec}+6)E_\mathrm{rec}/\hbar$. In the moving lattice frame, this emulates a force such that atoms are accelerated to larger values of the momentum. Once a desired final value is achieved, the Raman beams are extinguished, and in the following a Stern-Gerlach magnetic field is activated, with the gradient oriented orthogonally to the direction of the Raman beams. Following a free time of flight of \SI{12}{\milli\second}, subsequently, an absorption image is recorded, and by analysis of the deflection of atoms along and orthogonal to the axis of the Raman beams both the far field momentum distribution and the Zeeman state can be determined. From this data the magnetization $\langle m_J\rangle$ and the velocity can be obtained, as clear from noting that the expectation value for the kinetic momentum is $m\langle v_x\rangle = p_x- p_\mathrm{rec}\langle m_J\rangle$, from which one finds $\langle v_x\rangle=(\Delta\omega-\omega_z)/2k – (p_\mathrm{rec}/m)(6+\langle m_J\rangle)$.

\begin{figure}[htb]
	\includegraphics[width=\linewidth]{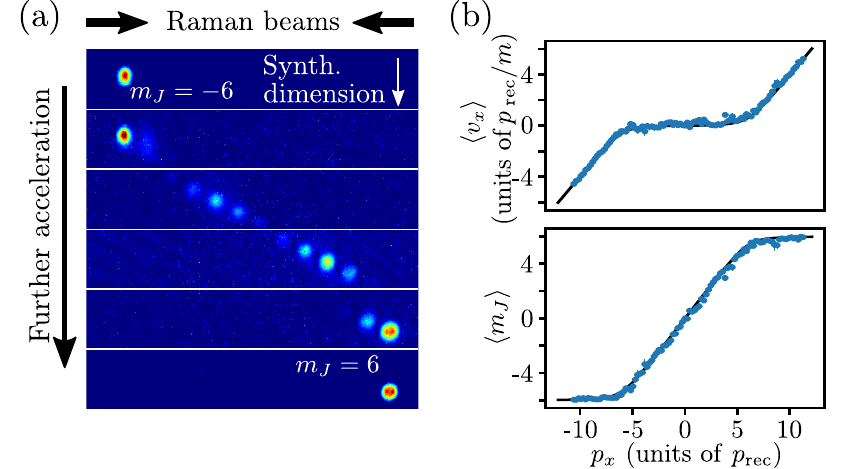}
	\caption{(a) Series of time of flight absorption images recorded for different acceleration times of atoms in the lowest energy band of the synthetic quantum Hall system. The visible deflection of atoms along the horizontal axis is due to the momentum transfer imparted from the Raman beams, while the deflection along the vertical scale stems from atoms experiencing a Stern-Gerlach force in an applied magnetic field gradient, allowing for analysis of the Zeeman state. The experimental data shows the temporal evolution following a Hall drift towards positive values of $m_J$. Both at the beginning and the end of the acceleration sequence, atoms populate an edge state (states $\ket{g_{-6},p_x+12\hbar k}$ and $\ket{g_6,p_x-12\hbar k}$ respectively), while in the bulk of the system the eigenstates are superpositions of several components. (b) The upper panel gives the variation of the measured atomic mean velocity (data points) on the canonical momentum $p_x$, along with theory for the group velocity (black line). The lower panel gives the corresponding variation of the mean value of the Zeeman quantum number $\langle m_J\rangle$ (data points) along with theory (black line). The used Rabi coupling is $\Omega \cong 1.44 E_{\mathrm{rec}}/\hbar$.}
	\label{meas:velo_mag} 
\end{figure}

Figure~\ref{meas:velo_mag}(a) shows experimental time-of-flight images recorded for different ramp times, with the axis of the Raman beams being indicated. The uppermost image was recorded at a large negative value of the canonical momentum $p_x$, at which the population is essentially purely in the $m_J=-6$ Zeeman level, corresponding to one of the edges of the topological system. For increased interaction times, atoms are further accelerated, upon which the Zeeman quantum number also increases, an issue that reflects the Hall drift in the synthetic dimension. In this central region, the population at each value of the canonical momentum $p_x$ is spread among several components. At even higher values of the momentum, the opposite edge state at $m_J=6$ is reached. Next, from such image sequences, we have determined the dependency of the atomic velocity $v_x$ on the canonical momentum $p_x$. Corresponding results are shown by the data points in the upper panel of Fig.~\ref{meas:velo_mag}(b), and the solid line for comparison gives the expected variation $v_{g,x}=dE/dp_x$, derived from the dispersion relation of the ground band. While on the edges the dispersion is linear, in the central region we clearly observe a nearly flat dispersion, as expected for the bulk of the topological quantum Hall system. Further, the bottom panel shows the variation of the mean value of the Zeeman quantum number $\langle m_J\rangle$ on the canonical momentum. Here clearly again, the edges of the system in state space along the synthetic dimension at $m_J=\pm J$ are visible, while in between, a smooth linear variation is seen. 

\begin{figure}
	\includegraphics[width=\linewidth]{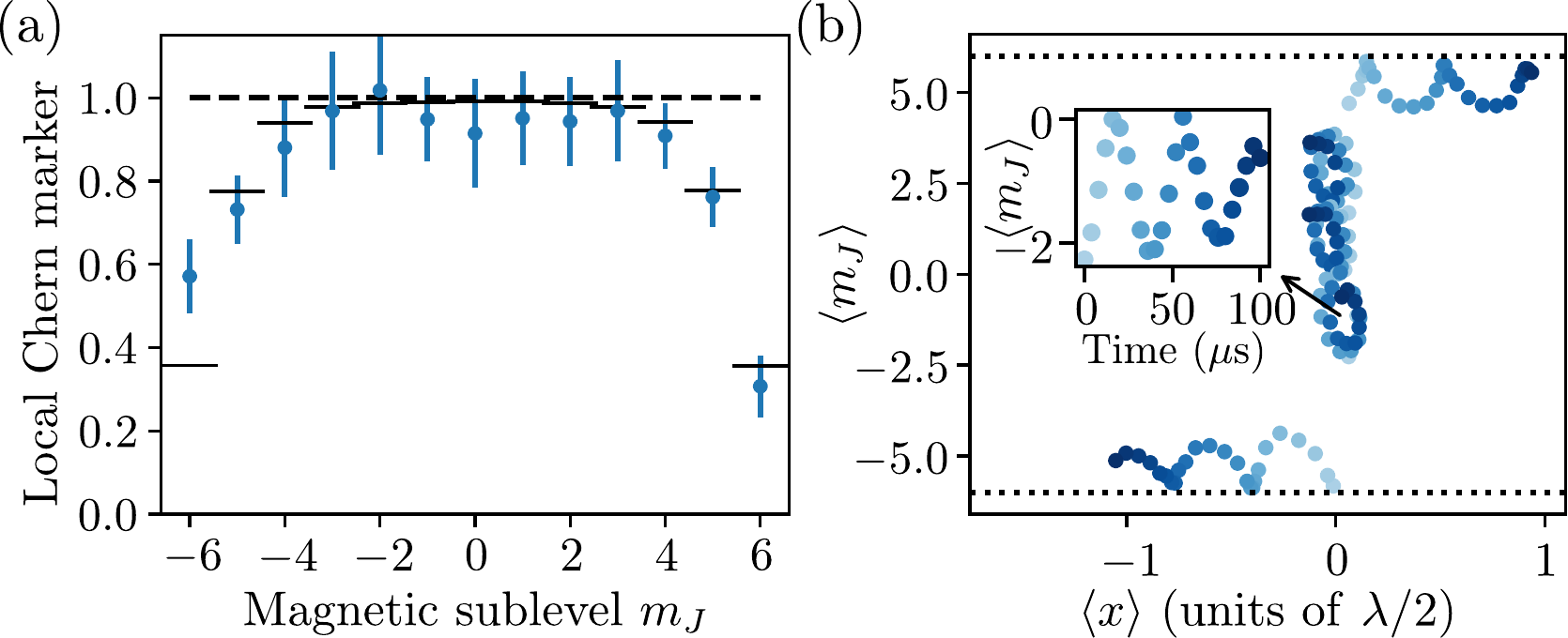}
	\caption{(a) Variation of the local Chern marker for the ground band (data points) along with theory (black lines) on the Zeeman quantum number. For an infinite system, one expects a value of unity of the Chern number, as is indicated by the visible dashed horizontal line. (b) Temporal evolution of atomic wavepackets along the real space dimension (horizontal axis) and synthetic dimension (vertical axis) following excitation to the next higher band by giving a velocity kick of $p_\mathrm{rec}/m$, for different initial conditions. While in the bulk of the system cyclotron orbits are observed, along the edges skipping orbits are seen. Each of the shown five different trajectories is the result of a separate measurement set performed with a specific value of the initially prepared momentum state. Temporal evolution is indicated by a change in color scale; see also the exemplary evolution along the synthetic dimension for one of the trajectories in the inset.
	}
	\label{meas:orbits}
\end{figure}

To characterize the topological properties of the system, we have determined the local Chern marker, a quantity introduced in Ref. \cite{bianco2011mapping}. The motivation here is that given the finite size of the system, the Chern quantum number, as a global quantity, cannot be defined. However, the local Chern number, which is given here by \cite{Chalopin2020}

\begin{equation}
	c(m_J) = \frac{1}{2\hbar k}\int dp_x \Pi_{m_J}(p_x)\frac{\partial}{\partial p_x}(p_x-m\langle v_x\rangle),
\end{equation}
well characterizes topology of the system on a local level. Here $\Pi_{m_J}(p_x)$ is a spin projection operator. Figure~\ref{meas:orbits}(a) gives the variation of the local Chern marker on the Zeeman quantum number. In the bulk of the system, as here defined by the region for which the measured group velocity is below 0.1 $p_\mathrm{rec}/m$, the average value of the local Chern number is $C=0.92 \pm 0.12$. Within experimental uncertainties, this well agrees with numerically expected value for our system of $C=0.98$, and is also close to the expectations for an infinite system of $C=1$. At the edges, the determined value of the local Chern number decreases, as expected when leaving the bulk. We attribute our corresponding data as evidence for a non-trivial topology of the system.

In further measurements, we have employed non-adiabatic transitions by providing a sudden velocity kick of $p_\mathrm{rec}/m$ to transfer part of the atomic wavepacket into the first excited band of the eigenenergy spectrum as to allow for a monitoring of the temporal evolution of superpositions of different Landau-like levels, and observe the characteristic system dynamics. Corresponding experimental data is given in Fig.~\ref{meas:orbits}(b), showing the temporal variation of the mean value of the real space position, as derived by temporally integrating the group velocity, and the magnetization, for different initial conditions. Temporal evolution is here indicated by a change in color scale, see also the exemplary diagram given in the inset. In the bulk of the synthetic Hall ribbon, we observe cyclotron orbits. The oscillation occur at frequency $\omega_c/2\pi\approx 25.3 \pm 0.3~\SI{}{\kilo\hertz} $ (see also the exemplary time trace in the inset), which represents our experimentally determined value for the cyclotron frequency. On the other hand, when atomic wavepackets are placed near the edge, skipping orbits directed in opposing directions in the top respectively bottom of the topological system are visible. In all cases, this well reflects observations in electron systems, as well as in other cold atom synthetic systems \cite{Mancini2015,stuhl2015visualizing,Chalopin2020}. The non-adiabatic mixing of eigenstates was for the measurements corresponding to the bulk region and upper edge performed by employing a non-adiabatic jump of the Raman beams frequency difference by $2 E_\mathrm{rec}/\hbar$, while for the lower edge Raman beams were turned on abruptly on two-photon resonance.

\begin{figure}
	\includegraphics[width=\linewidth]{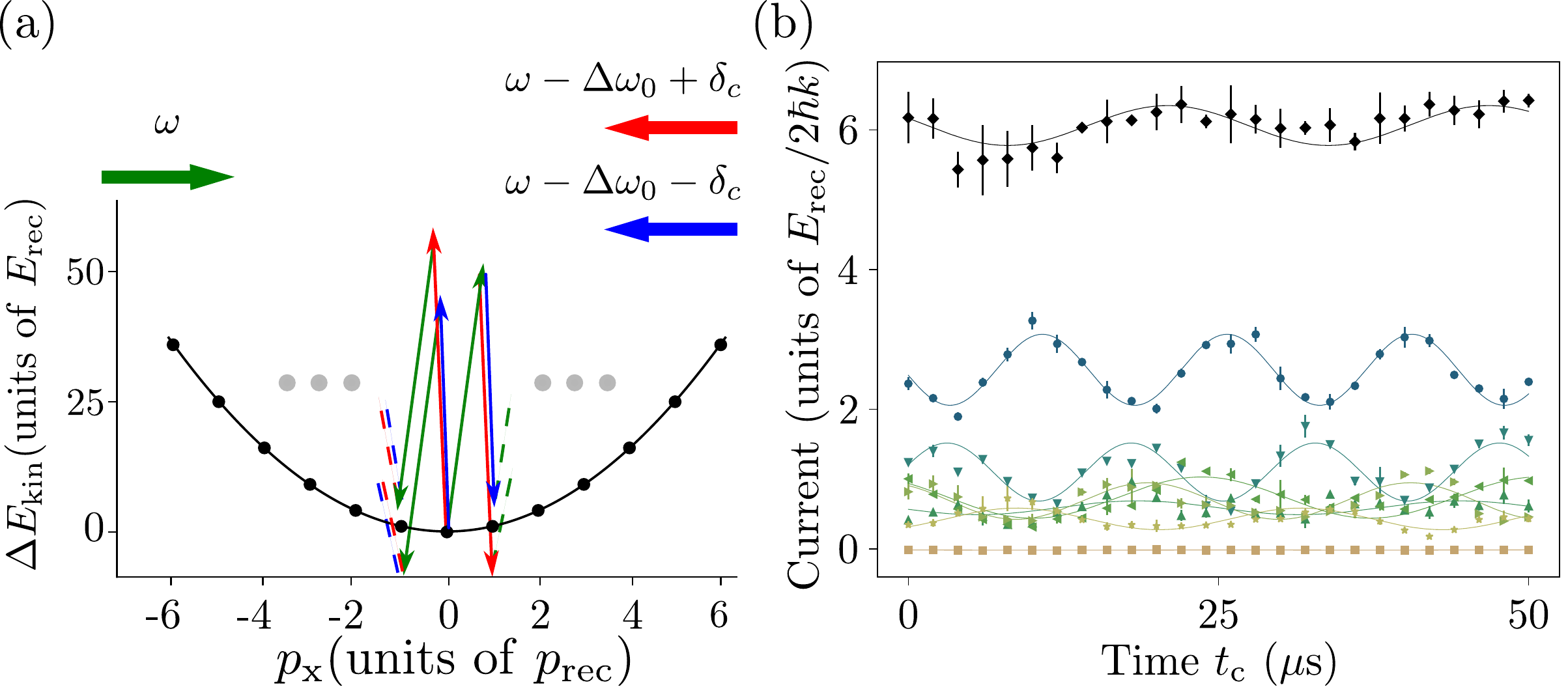}
	\caption{(a) Representation of the used bichromatic Raman scheme used to induce a chiral edge current in the synthetic system in an energy-momentum diagram. After preparing atoms in the lowest energy band at $p_x=0$, the bichromatic Raman drive is applied such that atoms are transferred to both edges simultaneously in the presence of the quadratic atomic dispersion. For the sake of clarity, the figure only shows the coupling combinations enabled by the bichromatic drive in the central region. (b) The black data points give the temporal variation of the chiral edge current of the synthetic quantum Hall system, as observed following a bichromatic pulse performed with the Raman beams transferring population to both edges of the system. The colored data points give contributions to the chiral current from individual pairs $(m_J, -m_J)$ of Zeeman levels: $m_J=\pm6$ (dots), $m_J=\pm5$ (triangles down), $m_J=\pm4$ (triangles up), $m_J=\pm3$ (triangles left), $m_J=\pm2$ (triangles right), $m_J=\pm1$ (stars), and the oranges squares give the (vanishing) contribution of the $m_J=0$ state to the chiral current. The solid lines are sinusoidal fits to the experimental data. }
	\label{meas:curr}
\end{figure}

In subsequent experiments, we have excited both edges of the quantum Hall ribbon simultaneously and determined the resulting chiral current. Other than earlier cold atom works studying chiral edge currents, which have been performed in synthetic Hall systems with three spin components \cite{Mancini2015,stuhl2015visualizing}, the here investigated atomic erbium system with its $\left(2J+1\right)=13$ Zeeman states has a large enough size of the synthetic dimension to exhibit distinct bulk and edge behavior. On the other hand, the large size of the synthetic system used here implies that a simultaneous transfer of population to both upper and lower edges requires a suitable state preparation scheme, which due to the large difference of associated momenta relates to the requirements in large area atom interferometry~\cite{Muller08}. In our experiment, we start by initially accelerating atoms adiabatically to a canonical momentum state of $p_x=0$, i.e., centered in the lowest band of the synthetic Hall system. Subsequently, we apply a $\SI{35}{\micro\second}$ long bichromatic pulse of the Raman beams performed with the two-photon detuning shifted by $\pm\delta_c, \mathrm{with}~\delta_c=\pm 7 E_\mathrm{rec}/\hbar$ respectively, as to induce a coupling of the central states to also the outer components $\ket{g_{\pm6}, p_x\mp 12\hbar k}$ by multiphoton Raman transitions in the presence of the quadratic atomic dispersion, see also the schematics of Fig.~\ref{meas:curr}(a). Subsequently, we return to a single frequency Raman drive (with two-photon detuning $\Delta\omega_0-\omega_z\cong12E_\mathrm{rec}/\hbar$, where $\Delta\omega_0\equiv\Delta\omega (p_x=0)$) such that a Landau-like flattened dispersion of the low energy bands is re-obtained and again $p_x=0$. This state preparation method leaves the population distributed among also highly excited levels of the band structure of the synthetic Hall system at $p_x=0$ [Fig.~\ref{fig:setup}(c)]. We find an essentially equal population of states propagating in opposite directions in the real space axis (x-axis) of the topological system. The chiral current of the synthetic Hall system is given by, 

\begin{equation}
	I_c = \sum_{m_J=1,...,6}n_{-m_J}v_{-m_J}- n_{m_J}v_{m_J},
\end{equation}
where $n_{\pm m_J}$ denotes the population of the respective Zeeman levels and $v_{\pm m_J}$ the associated velocities. For an experimental determination, the population distribution was determined by time of flight imaging at a variable holding time in the topological system after applying the above described state preparation sequence. The black diamonds in Fig.~\ref{meas:curr}(b) give the variation of the total chiral current versus time, and the colored data points show the contributions of Zeeman states of a given modulus of $m_J$ to the total chiral current. It is clearly visible that the dominant contributing to the chiral current is determined by the contributions of the outer Zeeman levels. The observed temporal modulation is attributed to a beating of the dressed state levels. Our experimental data shows that also the modulation contrast for the outermost Zeeman states is largest, specifically for $|m_J| = 5~ \mathrm{and}~6$ respectively, which form the dominant contribution to the four highest energetic dressed states near $ p_x=0$. While the splitting of avoided crossings of the higher energy bands - which is associated with the weak spin-orbit coupling between e.g. the $|g_6, p_x-12\hbar k\rangle$ and $|g_{-6}, p_x+12\hbar k\rangle$ outer components respectively - almost completely vanishes, a diagonalization of the coupled system shows that the two highest energetic dressed state levels at $p_x=0$ in addition to the dominating $m_J=\pm6$ components for the used Rabi frequency still have a $\sim$$10\%$ admixture of  $m_J=\pm5$, such that the presence of the visible temporal modulation is well understood. The observed oscillation frequency of the $|m_J|=5$ and $|m_J|=6$  contribution to the current of $67.2\pm 0.2$ \SI{}{\kilo\hertz} is clearly above the cyclotron frequency of low-lying bands, and within experimental uncertainties agrees with expected value for the size of the splitting of $66.3\pm1.3$ \SI{}{\kilo\hertz} for the used Rabi coupling of $\Omega\cong (1.35 \pm 0.21) E_\mathrm{rec}/\hbar$. A diagonalization of the band structure shows that for smaller Rabi couplings the splitting asymptotically approaches the difference of components in recoil energies in frequency units of $(6 p_\mathrm{rec})^2/2m – (5 p_\mathrm{rec})^2/2m = 11 E_\mathrm{rec}/h \cong \SI{73.8}{\kilo\hertz}$. Temporal modulations of frequency associated with the photon recoil have in earlier works been observed in multiple beam atomic interferometers and atom-based studies of the Talbot effect, where also coherences of interfering paths differing in momentum by integer multiples of the photon recoil are relevant \cite{Weitz97,Kumara88}.
	
To conclude, we have realized a synthetic quantum Hall system with an atomic erbium Bose-Einstein condensate. The non-trivial topological properties of the synthetic system have been verified for the flattened lowest energy bands by measuring the local Chern number, and furthermore closed and skipping orbits were observed by employing excitations between low-lying bands. In a study of the chiral edge dynamics the more complex, modulated structure of highly excited bands of the coupled system becomes relevant.

For the future, synthetic quantum Hall systems holds promise for the observation of fractional quantum Hall physics~\cite{Chalopin2020}. The non-vanishing angular momentum of the used rare-earth atomic species in the electronic ground state is expected to allow for a smaller heating from photon scattering than in alkali atoms~\cite{BenLev13}. This is a property shared with other rare earth or transition metal elements, as dysprosium, thulium, or titanium~\cite{Eustice20}. Specifically, elements with a not too high value of $J$, while certainly reducing the size of the synthetic dimension, have the benefit of a smaller dipole-dipole interaction, with a ratio of $\epsilon_\mathrm{dd}\cong0.48$ of dipole-dipole to the s-wave interaction for the erbium $^{168}$Er case (for dysprosium $^{162}$Dy: $\epsilon_\mathrm{dd} \cong 1.0$) when assuming the background scattering length \cite{Chomaz_2023}. Still, while the present experiment well operates at conditions of Bose-Einstein condensation, a reaching of the manybody regime is nevertheless, as to reduce dipolar loss, expected to require the use of small magnetic bias fields, as feasible with magnetic shielding, or the use of the fermionic isotope $^{167}\mathrm{Er}$ at high bias fields \cite{Benlev2015}. Other perspectives range from an observation of the dissipative response of the topological system \cite{Tran17,Tran18,Asteria2019} to the realization of Majorana-type physics \cite{Zhu11,Maghrebi15}.

We thank T. Chalopin and M. Fleischhauer for helpful discussions. Support by the DFG within the focused research center SFB/TR 185 (277625399) and the Cluster of Excellence ML4Q (EXC 2004/1 – 390534769) is acknowledged.

\section{Appendix}

\subsection{Spin-Orbit coupled Hamiltonian}
A detailed derivation of the spin-orbit coupled Hamiltonian of Eq.1 can be found in  Ref~\cite{chalopin2019quantum}, which is here summarized for sake of completeness. Using the Ansatz
\begin{equation}\label{h0}
	\hat{H} = \hat{H}_0 + \hat{H}_{\mathrm{int}},
\end{equation}
with $\hat{H}_0 = \hat{\mathbf{p}}^2/(2m) + \hbar\omega_z\hat{J}_z$ as bare Hamiltonian accounting for the kinetic and the Zeeman energies of an atom of mass $m$ and momentum $\mathbf{p}$ subject to an external magnetic field $B_z$ directed along the z-axis. Here $\hbar\omega_z$ denotes the Zeeman splitting between neighbouring spin projections and $\hat{J}_z$ the z-component of $\vec{J}=(\hat{J}_x,\hat{J}_y,\hat{J}_z)$, the dimensionless spin operator, with $\langle \hat{J}\rangle=J=6$ for erbium. The second term accounts for the atoms-light interaction, which can be written as~\cite{LeKien2013}
\begin{eqnarray}\label{eq:HamInt}
	\hat{H}_{\mathrm{int}} &=& V_0\bigg\{\alpha_s\left|\mathbf{u}\right|^2\mathbbm{1}-i\alpha_v(\mathbf{u}^*\times\mathbf{u})\frac{\hat{\mathbf{J}}}{2J}\\
	&+&\alpha_t\frac{3[(\mathbf{u}^*\hat{\mathbf{J}})(\mathbf{u}\hat{\mathbf{J}})+(\mathbf{u}\hat{\mathbf{J}})(\mathbf{u}^*\hat{\mathbf{J}})]-2|\mathbf{u}|^2\hat{\mathbf{J}}^2}{2J(2J-1)}\bigg\},
\end{eqnarray}
where $V_0 = 3\pi c^2\Gamma I/2\omega_0^3\Delta$, and $\alpha_s,\alpha_v$ and $\alpha_t$ the scalar, vectorial and tensorial polarizabilities, respectively. Further, $\mathbf{u}$ describes the position dependent polarisation vector, which for the presented crossed linear polarization configuration can be written as
\begin{equation}\label{eq:PolVec}
	\mathbf{u} = \frac{1}{2}e^{-i\omega_0t}\left[e^{ikx}(\hat{\mathbf{y}}+\hat{\mathbf{z}})+e^{-i(kx+(\omega_z+\delta)t)}(\hat{\mathbf{y}}-\hat{\mathbf{z}})\right].
\end{equation}
Here $\hat{\mathbf{y}}$ and  $\hat{\mathbf{z}}$ are corresponding unit vectors and $\delta$ describes an additional detuning. Using the definition for $\mathbf{u}$, the interaction term can be written as
\begin{eqnarray}
	\hat{H}_{\mathrm{int}} &=&\nonumber V_0\bigg[\alpha_s\hat{\mathbbm{1}}+\alpha_v\mathrm{sin}(\phi)\frac{\hat{J}_x}{2J}\\ &+&\alpha_t\frac{\hat{\mathbf{J}}^2-3\hat{J}_x^2+3\mathrm{cos}(\phi)\left(\hat{J}_y^2-\hat{J}_z^2\right)}{2J(2J-1)}\bigg],
\end{eqnarray}
where $\Phi = 2kz-(\omega_z+\delta)t$. We next apply two unitary transformations, one in the time domain given by
\begin{equation}
	\left|\Psi\right> = \hat{U}\left|\psi\right> = e^{i(\omega_z+\delta)t\hat{J}_z}\left|\psi\right>
\end{equation} 
followed by a rotating wave approximation, and one in position space given by
\begin{equation}
	\left|\Psi\right> = e^{i2k\hat{x}\hat{J}_z}\left|\psi\right>,
\end{equation}
so that the resulting expression can be written as a spin-orbit coupled Hamiltonian 
\begin{eqnarray}\label{eq:hamil:unitary}
	\hat{H}_{\mathrm{SOC}} &=&\nonumber \frac{\hat{p}_y^2 + \hat{p}_z^2}{2m}+\frac{(\hat{p}_x-2\hbar k\hat{J}_z)^2}{2m}-\hbar\delta\hat{J}_z\\ 
	&+& V_0\Bigg[\alpha_s\hat{\mathbbm{1}}-\frac{\alpha_v}{4J}\hat{J}_y+\frac{\alpha_t}{2J(2J-1)}\frac{3}{2}\hat{J}_z^2\Bigg].
\end{eqnarray}
Further, with $\alpha_v = \frac{90}{91},
\alpha_t = -\frac{22}{91}=-\frac{15}{2J+3}\frac{22}{91}$ \cite{chalopin2019quantum}, and neglecting the energy offset $\alpha_s\hat{\mathbbm{1}}$ as well as the trivial terms for the kinetic energies in the y and z directions, we have
\begin{equation}
	\hat{H}_{\mathrm{a}}= \frac{(\hat{p}_x-2\hbar k\hat{J}_z)^2}{2m}-\hbar\delta\hat{J}_z- \hbar\Omega\Bigg[\hat{J}_y+\frac{1}{2J+3}\hat{J}_z^2\Bigg],\label{eq:Ham:SOC}
\end{equation}
which equals Eq.1 of the main text.

\end{document}